\documentclass[twocolumn,showpacs,preprintnumbers,amsmath,amssymb]{revtex4}
\usepackage{graphicx}
\usepackage{epsfig}
\preprint{MgB2rapid}
\begin{document}

\title{Two-band parallel conductivity at terahertz frequencies in the superconducting state of MgB$_2$.}
\author{M. Ortolani$^1$, P. Dore$^2$, D. Di Castro$^2$, A. Perucchi$^{2,3}$, S. Lupi$^{2,3}$,\\ V. Ferrando$^4$, M. Putti$^4$, I. Pallecchi$^4$, C. Ferdeghini$^4$ and X. X. Xi$^5$}
\affiliation{$^1$CNR-Istituto di Fotonica e Nanotecnologie, Via Cineto Romano 42, I-00156 Rome, Italy}
\affiliation{$^2$ CNR-INFM Coherentia and Dipartimento di Fisica, Universit\`a di Roma La Sapienza, Piazzale Aldo Moro 2, I-00185 Rome, Italy}
\affiliation{$^3$Sincrotrone Trieste, Area Science Park, I-34012 Trieste, Italy}
\affiliation{$^4$CNR-LAMIA and Universit\`a di Genova, Via Dodecaneso 33, I-16146 Genoa, Italy}
\affiliation{$^5$Department of Physics Pennsylvania State University University Park, Pennsylvania 16802, USA}

\date{\today}

\begin{abstract}

The optical response  of the two-band superconductor MgB$_2$ has been studied in the 0.7-4 THz range on films with very low impurity level. The effect of the high-energy $\sigma$-gap is observed in the ratio $R_S/R_N$ between the normal and superconducting state reflectance, while in a neutron irradiated film with a slightly higher impurity level mainly the effect of the $\pi$-gap is evident as reported in previous experiments. At terahertz frequencies, the electrodynamic of MgB$_2$ can be well described by the two-band parallel conductivity model and is dominated by the $\pi$-bands when the impurity level is only slightly higher than that of an ultra-clean sample.

\end{abstract}

\maketitle

The two-band superconductivity of MgB$_2$ has raised a large interest and comprehensive theoretical and experimental studies of this material have been performed in the last years \cite{PhysC}. Thanks to the high $T_c$, the simple layered structure, the phononic origin of the superconducting pairing, and the amount of experimental and theoretical work already done, MgB$_2$ gives a unique opportunity for testing theoretical models of BCS superconductivity in the two-band case. Furthermore, MgB$_2$ epitaxial films are being extensively investigated for possible applications in superconducting devices and terahertz electronics \cite{Appl,Jin}. Infrared (IR) spectroscopy has largely contributed to the description of the physical properties of MgB$_2$, as discussed in detail in the recent comprehensive review of Kuzmenko \cite{Kuz}. In the normal state, this technique can indeed distinguish the different contributions of the two bands ($\sigma$ and $\pi$) to the frequency-dependent conductivity \cite{Kuz, Kak, Gur, Ddc}. In particular, reflectance measurements on the $ab$ plane can provide values of the two corresponding plasma frequencies $\Omega_i$ and scattering rates $\gamma_i=1/\tau_i~ (i=\sigma,\pi)$. In the superconducting state, several studies have been performed in the terahertz/far-IR range (here defined by photon energies below 16 meV, frequency $\omega<$ 130 cm$^{-1}$ or $\nu<4$ THz) since a mark of the superconducting gap $\Delta$ is expected in the reflectance or transmittance spectrum at $\hbar\omega \sim 2\Delta$ (optical gap) for an isotropic $s$-wave BCS superconductor. 

The most recent optical conductivity data from single crystals in the normal state show an overall consistency with band structure calculations \cite{Kuz,Massidda}. In the superconducting state, on the contrary, a two-gap description of the far-IR spectrum of MgB$_2$, based on independent BCS responses of $\sigma$ and $\pi$ bands, did fail in describing experimental data.
In fact, the far-IR response of the $ab$-plane of MgB$_2$ appears to be dominated by the $\pi$-band carriers \cite{Carr,Perucchi,Jung,Jin}. Indeed, no evidence of the high-energy gap $\Delta_{\sigma}$ was observed and a  feature was generally seen at $\hbar\omega\sim 2\Delta_\pi$ in the ratio $R_S/R_N$ ($T_S/T_N$), where $R_S$ ($T_S$) and $R_N$ ($T_N$) are the frequency-dependent reflectances (transmittances) in the superconducting and normal state, respectively. The BCS theory can be extended to the MgB$_2$ case by assuming a parallel sum of the conductivity of two independent bands with no cross terms, each conductivity being described by a BCS model generalized to arbitrary temperature and $\gamma_i$ values\cite{Ddc,zim} (for convenience, this model will be indicated hereafter as \emph{generalized BCS model}). However, in the far-IR response evaluated with this model the main feature shows up at $\hbar\omega\sim 2\Delta_\sigma$ when realistic values of the parameters $\Omega_i, \gamma_i$ and $\Delta_i$ are employed \cite{Kuz}. This inconsistency raises an important problem: while the dc transport properties and the dc-field penetration depth \cite{Golubov} can be described under the two-band parallel conductivity assumption, the present body of experimental evidence suggests that this assumption can be questioned when applied at terahertz  frequencies in the superconducting state.

\begin{figure}
{\hbox{\psfig{figure=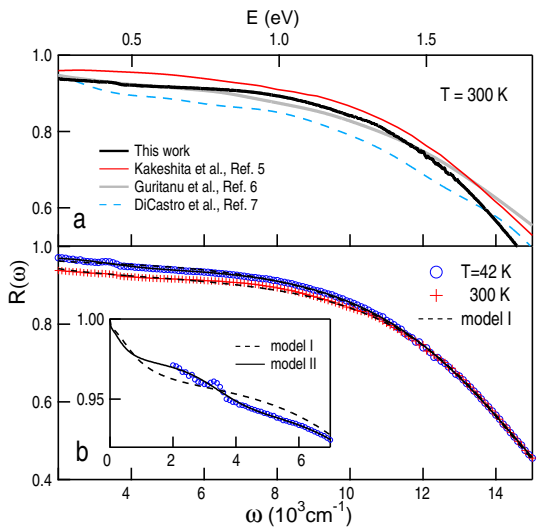,width=8cm}}}
\caption{(a) Infrared reflectivity $R(\omega)$ of MgB$_2$ films from the present work compared to that of crystals from Refs.\ 5,6 and 10. (b) Drude-Lorentz fits to $R(\omega)$ (see text).}
\end{figure}

The aim of the present work was to perform accurate measurements of the $R_S/R_N$ spectrum in the far-IR range in order to verify the validity of the two-band parallel conductivity assumption. Since the small size of high-quality MgB$_2$ single crystals makes low-temperature measurements in the far-IR range very challenging \cite{Perucchi,Ddc}, we used high-quality epitaxial films prepared by hybrid physical chemical vapor deposition, with $T_c$ = 41 K and optimal transport properties \cite{Xi}. The size of the film surface (5x5 mm), with an average roughness of few  nanometers, allows reflectance experiments in the whole infrared range, and the large film thickness ($d$ = 200 nm) strongly reduces the optical effect of the film substrate (see below). Since it is well known that MgB$_2$ samples quickly degrade in air in a few minutes \cite{Xi2}, each film was sealed under vacuum in a quartz ampoule after growth and all the optical experiments were continuously performed in a vacuum of 10$^{-6}$ mbar.

We performed FT-IR reflectance measurements at the infrared beamline SISSI at the synchrotron Elettra (Trieste, Italy) by using a He-flow cryostat, with a sapphire optical window and a cold finger which allows to position the pre-aligned sample and reference gold mirror on the incident beam with micrometric precision. We thus obtained the absolute reflectivity $R(\omega)$ from 2000 to 15000 cm$^{-1}$ at 300 K and 50 K. In the far-IR, by exploiting the transparency of the sapphire window, we used synchrotron radiation to keep the focal spot size smaller than the film size with no loss of signal-to-noise ratio. We measured the intensity reflected by the sample by cycling the temperature in the 5-50 K range, without collecting the reference spectrum. In this way one avoids any variation in the sample position and orientation, which can produce frequency-dependent systematic errors in $R(\omega)$. By using the 42 K data as the normal state reference, we obtained $R(T)/R(42 $K) spectra with an accuracy of $\pm 0.1$\%.

	The agreement among the room temperature $R(\omega)$ of the  film and of single-crystals \cite{Kak, Gur,Ddc} is good (see Fig.\ 1a), indicating that the optical quality of our film is equivalent to that of single crystals. The  $R(\omega)$ of the film measured at 50 K is compared with the one at 300 K in Fig.\ 1b. As expected, $R(\omega)$ increases with decreasing temperature owing to the decrease of the scattering rates. 
We first performed a fit of $R(\omega)$ at 300 K by employing the Drude-Lorentz model including two Drude terms associated to the $\sigma$ and $\pi$ bands, and a Lorentz contribution centered around 20000 cm$^{-1}$ representing an optically active $\sigma \to \pi$ interband transition (model I) \cite{Kak,Gur}. We obtained an acceptable description of $R(\omega)$, as shown in Fig. 1b, with $\Omega_\sigma = 3.8$ eV, $\gamma_\sigma = 50$ meV, $\Omega_\pi = 5$ eV, and $\gamma_\pi = 370$ meV (the uncertainty on these values is of the order of 5\%). These parameter values are consistent with previous ones \cite{Kak, Gur}, as expected on the basis of the consistency among the different reflectivity spectra (see Fig.\ 1). We note that the $\gamma_i$ values at 300 K should be regarded as effective, model dependent values. Indeed, in a system like MgB$_2$, characterized by a strong electron-phonon coupling, the simple Drude model is inadequate at mid-IR frequencies. The extended Drude model \cite{Gur}, in which both scattering rate and effective mass are frequency dependent, should be applied instead. However, the generalization of the latter model to the two-band case is well beyond the scopes of the present paper.

At 50 K, by using the same values of the plasma frequencies, the agreement between the measured spectrum and best-fit curve is not satisfactory below 6000 cm$^{-1}$, as shown in the inset of Fig.\ 1b. The agreement with data is significantly improved if a mid-IR Lorentz oscillator at $0.4$ eV is included (model II). Although this oscillator may be partially due to the $\sigma \to \sigma$ interband excitation observed in a previous single-crystal measurement  \cite{Ddc}, it should be mainly regarded as an effective tool to compensate the above mentioned inadequacy of model I at mid-IR frequencies. Therefore, the 50 K best-fit scattering rates provided by model II ($\gamma_\sigma = 11$ meV, $\gamma_\pi = 75$ meV) do correctly describe the conductivity at far-IR frequencies, where the effect of the $0.4$ eV oscillator is negligible. We remark that these $\gamma_\sigma$ and $\gamma_\pi$ values are only determined by intrinsic impurities owing to the vanishing phonon population at low temperature in MgB$_2$ \cite{Masui,Marina} and are close to those previously reported for a single crystal ($\gamma_{\sigma,imp}=12.4$ meV, $\gamma_{\pi,imp}=85.6$ meV  \cite{Gur}).

As to far-IR results, the $R(T)/R(42$ K) measured at 50 and 5 K are shown in Fig.\ 2a. For $T>T_c$, as expected, the $R(T)/R(42 $K) curve is flat within experimental uncertainties, while on decreasing temperature below $T_c$ the $R(T)/R(42 $K) spectrum increases and a clear edge structure around 14 meV becomes evident at $T$ = 5 K. We remark that the high accuracy of the present synchrotron measurements allows the detection of an effect as small as $\sim$ 0.5\%. To discuss this result, we calculated the $R_S/R_N$ ratio by using the generalized BCS model introduced above, to be compared to $R$(5 K)$/R$(42 K). As input parameters, we employed the $\Omega_i$ and $\gamma_i$ values at 50 K above reported, and the $\Delta_i$ values obtained by tunneling spectroscopy on films of the same kind ($2\Delta_\sigma = 14.4$ meV, $2\Delta_\pi = 4.6$ meV \cite{Iavarone}). The $R_S/R_N$ spectra obtained by considering both a finite film thickness ($d$ = 200 nm) and a semi-infinite medium \cite{Kuz} are not appreciably different, as shown in Fig.\ 2a. By remarking that the model spectra are not best-fit to the data but results of calculations with no free parameters, the resemblance with the data is highly significant. In particular, an edge structure around 14 meV due to the $\sigma$ gap is observed in both experimental and model spectra, while in previous works it was concluded that the far-IR response is dominated by the $\pi$ bands since an edge was observed between 4 and 7 meV in all cases (see Refs.\ \onlinecite{Carr,Perucchi}, and Ref.\ \onlinecite{Kuz} for a review). The observation of the 14 meV edge is a crucial result, since it finally allows to reconcile the far-IR observation with the predictions of BCS calculations. We remark that this result has been obtained on a film with very low impurity level\cite{Xi} and minimal exposure to the atmosphere. We will thus refer to this sample as \emph{ultra-clean} film.

\begin{figure}
{\hbox{\psfig{figure=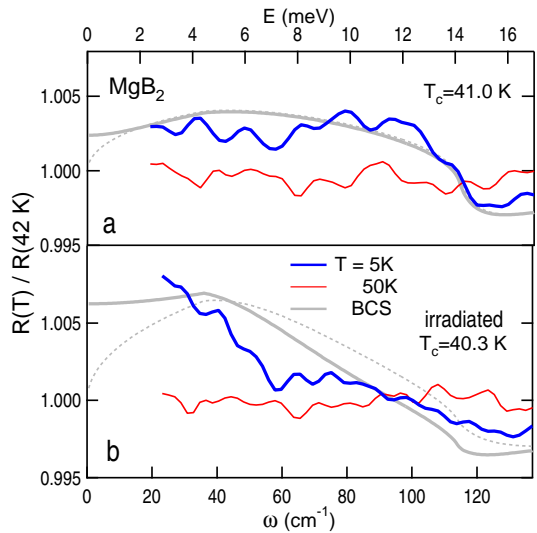,width=8cm}}}
\caption{Far-infrared reflectivity ratio (thick blue line) compared to BCS calculations (thick grey line: film thickness $d$ = 200 nm, dotted line: semi-infinite medium) with $\Omega_\sigma = 3.8$ eV, $\gamma_\sigma = 11$ meV ($\gamma_\sigma = 37$ meV in (b)), $\Delta_\sigma = 7.2$ meV, $\Omega_\pi = 5$ eV, $\gamma_\pi = 75$ meV, $\Delta_\pi = 2.3$ meV.)}
\end{figure}

To investigate the effect of the impurity level, we performed far-IR measurements on a second film which was previously irradiated with thermal neutrons (fluence of 10$^{16}$ particles/cm). Preparation and physical properties of irradiated films are fully described in Refs.\ \onlinecite{Marina,Pallecchi,Ferrando}. A high fluence of 10$^{19}$ particles/cm is known to suppress superconductivity in MgB$_2$ because of the production of a huge amount of lattice defects \cite{Canfield}. On the contrary, in the case of a low-fluence irradiated film like the one we investigated, the density of lattice defects is too small to modify the band structure or the electron-phonon coupling, and only an increase of the impurity scattering rates is observed, as shown by residual resistivity $\rho_0$ \cite{Ferrando}, Andreev reflection \cite{Gonnelli}, critical field and specific heat \cite{Marina} measurements. The measured reflectivity ratio $R(T)/R(42 $K) is shown in Fig. 2b for $T$=50 K and 5 K. A clear edge is evident around 7 meV, i.\ e.\ around 2$\Delta_\pi$, while a much smoother increase close to the noise level is observed below 12 meV.

In a film from the same batch as the one we investigated and irradiated with the same fluence (sample IRR15 in Ref. \onlinecite{Ferrando}), the $T_c = 40.3$ K and the gap values measured by tunnelling spectroscopy \cite{Dicapua} are very close to those of unirradiated films, while $\rho_0$ increases by a factor $\simeq$ 2 \cite{Ferrando}. This effect can be mainly attributed to an increase of $\gamma_\sigma$, as shown by a detailed analysis of magnetoresistance data \cite{Pallecchi} and by recent first-principle computations \cite{Monni}. In a reasonable approximation, the $R_S/R_N$ ratio can be evaluated within the generalized BCS model by using the same parameters of the ultra-clean film case, with the exception of $\gamma_\sigma$ increased from 11 to 37 meV. This value matches $\gamma_\pi/\gamma_\sigma = 2.0$ evaluated for the IRR15 sample in Ref.\ \onlinecite{Monni} and it is also compatible with the above cited increase of $\rho_0$. Remarkably, a reasonable agreement is obtained among the experimental data and the model spectra in Fig.\ 2b evaluated for both the $d$ = 200 nm film and a semi-infinite medium in the well accessible spectral range, i.\ e.\ above 4 meV. This result indicates that, when the scattering rates are slightly higher than those of the ultra-clean film, the far-IR spectrum appears to be dominated by the $\pi$-band carriers, as previously observed in a number of MgB$_2$ samples \cite{Carr,Jung,Perucchi,Ddc}. 

\begin{figure}
{\hbox{\psfig{figure=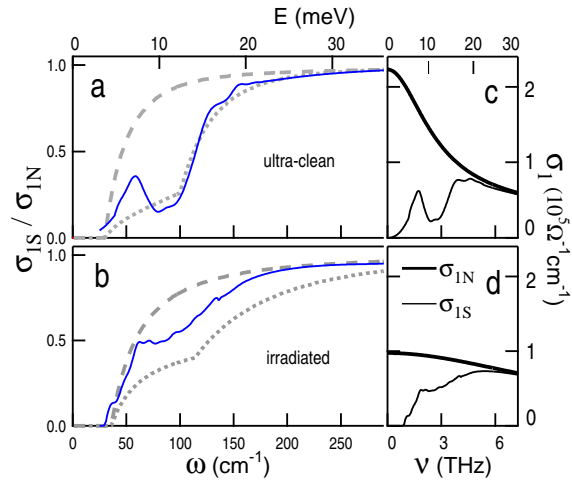,width=8cm}}}
\caption{(a,b) Continuous lines: optical conductivity change across the transition of the two MgB$_2$ films  compared to BCS calculations with parameters as in Fig.\ 2 (grey dots, two-gap) and with $\Delta_\sigma=\Delta_\pi = 2.3$ meV (grey dashes, one-gap). (c,d) Conductivity evaluated trough the Drude model (normal state, thick lines) and Kramers-Kronig transformations (superconducting state, thin lines). 1 THz corresponds to 33 cm$^{-1}$ or $4.1$ meV.}
\end{figure}

For a more straightforward interpretation of the far-IR data, we now discuss the obtained results in terms of the real part of the optical conductivity $\sigma_1(\omega)$. For the ultra-clean film, the normal-state $\sigma_{1N}$ was obtained through Kramers-Kronig transformation (KK) of $R_N$ given by the 50 K data in Fig.\ 1b extrapolated with the best-fit profile (model II). In the superconducting state, $\sigma_{1S}$ was calculated through KK of $R_S = R_N \cdot R$(5 K)$/R$(42 K). The ratio $\sigma_{1S}/\sigma_{1N}$ can be regarded as a well approximated experimental quantity, since most systematic uncertainties due to the KK do cancel out in the ratio \cite{Perucchi}, and also highlights the optical gaps, since it equals the transition probability variation across the superconducting transition. The ratio $\sigma_{1S}/\sigma_{1N}$ of the ultra-clean film reported in Fig.\ 3a exhibits two gaps, one at 6 meV and the other at 14 meV, the latter being more evident, to be associated to the $\pi-$ and $\sigma$-gap respectively. The same quantity obtained from the two-band generalized BCS model closely tracks the data, although the peak structure at 8 meV is not reproduced. On the contrary, the prediction for the one-gap case, obtained by equaling $\Delta_\sigma$ to $\Delta_\pi$ = 2.3 meV and leaving all other parameters unchanged, is not at all consistent with our result (see Fig.\ 3a).

Finally, we evaluate the optical constants of the two films by simply considering a parallel sum of two Drude terms defined by the parameters reported in the caption of Fig.\ 2. For the irradiated film, $\sigma_{1S}$ was also obtained through KK of $R_S = R_N \cdot R$(5 K)$/R$(42 K). The overall shape of $\sigma_{1S}/\sigma_{1N}$, shown in Fig.\ 3b, is in qualitative agreement with the prediction of the two-gap generalized BCS model. Therein, the most evident feature is a gap at 6 meV ($\pi$-gap), while the gap at 14 meV ($\sigma$-gap) is much less evident. The inspection of the separate $\sigma_{1N}$ and $\sigma_{1S}$ shown in Fig.\ 3c-d may elucidate the difference between the two films. $\sigma_{1N}$ of the ultra-clean film is dominated by the narrow Drude contribution due to $\sigma$ bands and decreases abruptly with frequency (see Fig.\ 3c), hence a steeper absorption edge in $\sigma_{1S}/\sigma_{1N}$ is observed around $2\Delta_\sigma$=14 meV in Fig.\ 3a. For the irradiated film with higher impurity level, $\sigma_{1N}$ decreases more slowly with $\omega$ owing to a larger $\gamma_\sigma$ (see Fig.\ 3d), then a clear absorption edge in $\sigma_{1S}/\sigma_{1N}$ appears around 5-7 meV in Fig.\ 3b. The latter situation is similar to what previously observed in both MgB$_2$ films \cite{Jin,Carr,Jung,Pronin} and single crystals \cite{Perucchi,Ddc}. This analysis can explain why the effect of the high-energy gap due to the $\sigma$ bands can be unambiguously observed only in samples with a very low impurity level, and shows that the two-band parallel conductivity model describes fairly well the terahertz conductivity of MgB$_2$ in both the normal and the superconducting states.

In conclusion, we performed IR reflectivity measurements on high-quality MgB$_2$ films. We find a close correspondence between our far-IR $R_S/R_N$ spectrum and that predicted by a generalized BCS model assuming a parallel sum of the conductivity of two independent bands. Far-IR measurements on a film with a slightly higher impurity level lead to a $R_S/R_N$ spectrum in which only the effect of the low-energy $\pi$ gap is well evident, thus explaining the inconsistency between theory and previous far-IR/terahertz experiments in terms of different impurity levels. Of general interest is the proof that the parallel sum of the conductivity of two superconducting bands can well describe the electrodynamic response of MgB$_2$ at terahertz frequencies.

The work at Penn State is supported in part by NSF under Grant No. DMR-0306746 and by ONR under Grant No. N00014-00-1-0294. M.O. acknowledges support from BESSY and IB-Berlin.


\begin{thebibliography}{}
\bibitem{PhysC}
S. Tajima, I. Mazin, D. van der Marel and H. Kumakura, in \textit{Recent Advances in MgB$_2$ Research}, Phys. C  \textbf{456}, 1 (2007).
\bibitem{Appl}
S. Cherednichenko, V. Drakinskiy, K. Ueda, and M. Naito, Appl. Phys. Lett.  \textbf{90}, 023507 (2007); Ke Chen, Y. Cui, Qi Li, X.X. Xi, Shane A. Cybart, R.C. Dynes, X. Weng, E.C. Dickey, and J.M. Redwing, Appl. Phys. Lett. \textbf{88}, 222511 (2006).
\bibitem{Jin}
B.B. Jin, P. Kuzel, F. Kadlec, T. Dahm, J.M. Redwing, A.V. Pogrebnyakov, X.X. Xi, and N. Klein, Appl. Phys. Lett. \textbf{87}, 092503 (2005).
\bibitem{Kuz}
A.B. Kuzmenko, Phys. C  \textbf{456}, 63 (2007).
\bibitem{Kak}
T. Kakeshita, S. Lee, and S. Tajima, Phys. Rev. Lett. \textbf{97}, 037002 (2006).
\bibitem{Gur}
V. Guritanu, A.B. Kuzmenko, D. van der Marel, S.M. Kazakov, N.D. Zhigadlo, and J. Karpinski, Phys. Rev. B \textbf{73}, 104509 (2006).
\bibitem{Ddc}
D. Di Castro, M. Ortolani, E. Cappelluti, U. Schade, N.D. Zhigadlo, and J. Karpinski, Phys. Rev. B \textbf{73}, 174509 (2006).
\bibitem{Massidda}
G. Satta, G. Profeta, F. Bernardini, A. Continenza, and S. Massidda Phys. Rev. B  \textbf{64}, 104507 (2001)
\bibitem{Carr}
J.J. Tu, G.L. Carr, V. Perebeinos, C.C. Homes, M. Strongin, P.B. Allen, W.N. Kang, E.M. Choi, H.J. Kim, and S.I. Lee, Phys. Rev. Lett. \textbf{87}, 277001 (2001).
\bibitem{Perucchi}
A. Perucchi, L. Degiorgi, J. Jun, M. Angst, and J. Karpinski, Phys. Rev. Lett. \textbf{89}, 097001 (2002)
\bibitem{Jung}
J.H. Jung, K.W. Kim, H.J. Lee, M.W. Kim, T.W. Noh, W.N. Kang, H.J. Kim, E.M. Choi, C.U. Jung, S.I. Lee, Phys. Rev. B \textbf{65}, 052413 (2002).
\bibitem{zim}
W. Zimmermann, E.H. Brandt, M. Bauer, E. Seider, and L. Genzel, Phys. C \textbf{183}, 99 (1991).
\bibitem{Golubov}
A.A. Golubov, A. Brinkman, O.V. Dolgov, J. Kortus, and O. Jepsen, Phys. Rev. B \textbf{66}, 054524 (2002).
\bibitem{Xi}
A.V. Pogrebnyakov, J.M. Redwing, S. Raghavan, V. Vaithyanathan, D.G. Schlom, S.Y. Xu, Q. Li, D.A. Tenne, A. Soukiassian, X.X. Xi, M.D. Johannes, D. Kasinathan, W.E. Pickett, J.S. Wu, and J.C. Spence, Phys. Rev. Lett. \textbf{93}, 147006 (2004).
\bibitem{Xi2}
Y. Cui, J. E. Jones, A. Beckley, R. Donovan, D. Lishego, E. Maertz, A. V. Pogrebnyakov, P. Orgiani, J. M. 
Redwing, and X. X. Xi, IEEE Trans. on Appl. Supercond. \textbf{15}, 224 (2005).
\bibitem{Masui}
T. Masui, Phys. C  \textbf{456}, 102 (2007).
\bibitem{Marina} 
 M. Putti, M. Affronte, C. Ferdeghini, P. Manfrinetti, C. Tarantini, and E. Lehmann, Phys. Rev. Lett. \textbf{96}, 077003 (2006).
\bibitem{Iavarone}
M. Iavarone, R. Di Capua, A.E. Koshelev, W.K. Kwok, F. Chiarella, R. Vaglio, W.N. Kang, E.M. Choi, H.J. Kim, S.I. Lee, A.V. Pogrebnyakov, J.M. Redwing, and X.X. Xi, Phys. Rev. B \textbf{71}, 214502 (2005).
\bibitem{Pallecchi}
I. Pallecchi, V. Ferrando, E. Galleani D'Agliano, D. Marre, M. Monni, M. Putti, C. Tarantini, F. Gatti, H.U. Aebersold, E. Lehmann, X.X. Xi, E.G. Haanappel, and C. Ferdeghini,  Phys. Rev. B \textbf{72}, 184512 (2005).
\bibitem{Ferrando}
V. Ferrando, I. Pallecchi, C. Tarantini, D. Marre, M. Putti, C. Ferdeghini, F. Gatti, H.U. Aebersold and E. Lehmann, J. Appl. Phys. \textbf{101}, 043903 (2007).
\bibitem{Canfield}
R.H.T. Wilke, S.L. Bud'ko, P.C. Canfield, J. Farmer, and S.T. Hannahs, Phys. Rev. B \textbf{73}, 134512 (2006).
\bibitem{Gonnelli}
D. Daghero, A. Calzolari, G.A. Ummarino, M. Tortello, R.S. Gonnelli, V.A. Stepanov, C. Tarantini, P. Manfrinetti, and E. Lehmann, Phys. Rev. B \textbf{74}, 174519 (2006).
\bibitem{Dicapua}
R. Di Capua, H.U. Aebersold, C. Ferdeghini, V. Ferrando, P. Orgiani, M. Putti, M. Salluzzo, R. Vaglio, and X.X. Xi, Phys. Rev. B \textbf{75}, 014515 (2007).
\bibitem{Monni}
M. Monni, I. Pallecchi, C. Ferdeghini, V. Ferrando, A. Floris, E. Galleani d'Agliano,
E. Lehmann, I. Sheikin, C. Tarantini, X.X. Xi, S. Massidda, M. Putti, unpublished, cond-mat/07063591.
\bibitem{Pronin}
A.V. Pronin, A. Pimenov, A. Loidl, and S. I. Krasnosvobodtsev, Phys. Rev. Lett. \textbf{87}, 097003 (2001).
\end{thebibliography}
\end{document}